\documentclass[twocolumn,showpacs,preprintnumbers,amsmath,amssymb]{revtex4}

\usepackage{graphicx}
\usepackage{dcolumn}
\usepackage{bm}

\begin{document}


\title{Ultrahigh energy cosmic ray acceleration in newly born magnetars \\and their associated gravitational wave signatures}

\author{Kumiko Kotera}
\email{kotera@uchicago.edu} 
\affiliation{
Department of Astronomy \& Astrophysics, Enrico Fermi
  Institute, and Kavli Institute for Cosmological Physics, The
  University of Chicago, Chicago, Illinois 60637, USA.}

\date{\today}

\begin{abstract}
Newly born magnetars are good candidate sources of ultrahigh energy cosmic rays. These objects can in principle easily accelerate particles to the highest energies required to satisfy the ultrahigh energy cosmic ray scenario ($E\sim10^{20-21}$~eV), thanks to their important rotational and magnetic energy reservoirs. Their acceleration mechanism, based on unipolar induction, predicts however a hard particle injection that does not fit the observed ultrahigh energy cosmic ray spectrum. Here we show that an adequate distribution of initial voltages among magnetar winds can be found to soften the spectrum. We discuss the effect of these distributions for the stochastic gravitational wave background signature produced by magnetars. The magnetar population characteristics needed to fit the ultrahigh energy cosmic ray spectrum could lead in most optimistic cases to gravitational wave background signals enhanced of up to four orders of magnitudes in the range of frequency $1-100$~Hz, compared to the standard predictions. These signals could reach the sensitivities of future detectors such as DECIGO or BBO. 
\end{abstract}

\pacs{Valid PACS appear here}
                             
                             
\maketitle

\section{Introduction}
The origin of ultrahigh energy cosmic rays (UHECRs) is still unknown (see \cite{KO11} for a review). From the point of view of simple particle confinement energetics (the so-called ``Hillas criterion" \cite{Hillas84}), young millisecond magnetars are one of the most promising candidate sources. Magnetars are isolated neutron stars with extremely strong surface dipole fields of order $B_{\rm d}\sim 10^{15}$~G and fast rotation at birth with initial rotation period $P_{\rm i}\sim 1$~ms (see \cite{Woods06,Harding06,Mereghetti08} for reviews). Because the source of energy for their radiative emission is magnetism, their dissipative properties are distinct from those of radio pulsars. Their existence was postulated by Ref.~\cite{Duncan92} and they are accepted as a plausible explanation for Soft Gamma Repeaters and Anomalous X-ray Pulsars (e.g., \cite{Kouveliotou98,Kouveliotou99,Baring01}). 

Ordinary pulsars have long been discussed as good candidates to accelerate charged ions through unipolar induction mechanism (see, e.g., \cite{Shapiro83} and references therein). Rapidly rotating neutron stars generally create relativistic outflows (``winds"), where the combination of the rotational energy and the strong magnetic field induces an electric field ${\bf E} =-{\bf v}\times {\bf B}/c$ (where {\bf v} and {\bf B} are the velocity and the magnetic field of the outflowing plasma). The wind thus presents voltage drops where charged particles can be accelerated to high energy. Ordinary pulsars however do not supply enough energy to reach the highest energies ($E>10^{20}$~eV). 

Magnetars on the other hand possess important rotational and magnetic energy reservoirs at birth that should enable them to accelerate easily, in principle, particles to $E>10^{20}$~eV \cite{Blasi00,Arons03}. They were introduced as possible progenitors of ultrahigh energy cosmic rays (UHECRs) during the ``AGASA era". If cosmic-rays originate from cosmological distances, their flux at the highest energies ($E\gtrsim 6\times 10^{19}$~eV) should be suppressed due to interactions with the cosmological background photons, creating a feature in the spectrum called the ``GZK cut-off" \cite{G66,ZK66}. The fact that the AGASA experiment did not observe such a feature \cite{Takeda98} led Ref.~\cite{Blasi00} to develop a model of acceleration of ultrahigh energy iron nuclei in young strongly magnetized Galactic neutron star winds. Ref.~\cite{Arons03} followed the same trend, but suggested that the hard injection spectrum produced by each magnetar could account for the absence of GZK cut-off, even with an extragalactic magnetar population scenario. 

The latest experiments report however that a suppression reminiscent of the GZK cut-off is present at the highest energy end of the UHECR spectrum \cite{Abbasi08,Abraham10}. The cosmic ray spectrum observed by the Pierre Auger Observatory can be described as a broken power-law, $E^{-\lambda}$, with spectral index $\lambda\sim 3.3$ below the break (called ``ankle") around $10^{18.6}$~eV, and $\lambda \sim 2.6$ above, followed by a flux suppression above $\sim10^{19.5}$~eV \cite{Abraham10}. The mainstream UHECR models view the ankle region as a transition between cosmic rays produced by Galactic and extragalactic source populations. Given this situation, the hard injection spectral index of $\lambda=1$ produced in the above magnetar scenarios is no longer an advantage. Figure~\ref{fig:crspec_uni} shows indeed that it is challenging to fit the observed spectrum down to the ankle energy with extragalactic sources injecting such a hard spectrum. The introduction of various source emissivity evolutions (described in Section~\ref{section:integrated}) are insufficient to reconcile the calculated and observed spectra, even for strong evolution cases. 

In this paper, we show that an adequate distribution of initial voltages among extragalactic magnetar winds can be found to soften the overall UHECR spectrum. These distributions result in some cases in tighter constraints on the magnetar population rate that is required to account for the observed UHECR flux, than suggested in Ref.~\cite{Arons03}. 

The magnetar model envisages that a wound-up, mainly toroidal magnetic field with strength $B_{\rm t}>10^{15}$~G characterizes the neutron star interior \cite{Thompson93}. Such strong internal fields should lead to a substantial deformation of the neutron star and thus to the emission of gravitational waves, provided that the magnetic distortion axis and the rotation axis of the star are not aligned \cite{Bonazzola96,Konno00}. Young millisecond magnetars should thus be strong gravitational wave emitters \cite{Palomba01,Cutler02,Stella05}. 

Gravitational wave signals from individual magnetars might be detected by instruments of the generation of Advanced LIGO \cite{Stella05}, and reveal properties about their magnetic field and initial rotation period that are crucial to probe magnetars as UHECR accelerators. However, such signals should not be observed in coincidence with UHECR events. Indeed, the time delay that cosmic rays experience by magnetic deflection during their propagation in the intergalactic medium, relative to the gravitational waves going in geodesics, is more than several hundreds of years.

Another gravitational wave signature that one can seek is the stochastic background emitted by the ensemble of magnetars. The signal calculated up to now assuming a population of magnetars with identical properties, were weak especially in the frequency range that should be observed by future experiments such as BBO or DECIGO \cite{Regimbau08}. We demonstrate in this paper that the distributions of magnetar characteristics required to account for the observed UHECR spectrum could, in some cases, significantly enhance the signal level in the frequency range $1-100$~Hz. If such signatures were detected, they could help determine if magnetars are indeed capable of accelerating the highest energy particles in the Universe. 

The layout of this paper is as follows. In Section~\ref{section:energetics}, we provide a synthetic picture of the production of UHECRs by newly-born magnetars, not entering into the detailed modeling of the acceleration mechanism or the escape of particles. In Section~\ref{section:integrated}, we calculate the UHECR flux obtained for various magnetar distributions and discuss the magnetar occurrence rate necessary to account for the observed cosmic ray flux. We calculate the implication of such distributions on the stochastic gravitational wave background in Section~\ref{section:gw}. Our results are further discussed in Section~\ref{section:discussion}.

\section{Magnetar energetics for UHECR acceleration}\label{section:energetics}

In this section, we provide a synthetic picture of the production of UHECRs by newly-born magnetars, following the work of Refs.~\cite{Blasi00,Arons03}. These authors propose that UHECRs are accelerated in the relativistic wind of rapidly spinning and strongly magnetized magnetars by unipolar induction. The toy model described below needs to be further investigated on several issues, for instance on the nature of the ions injected in the wind, on the mechanism through which the current in the wind taps the available voltage (i.e., the actual acceleration mechanism), and on the escape of the accelerated particles from the wind, and the surrounding supernova envelope \cite{Fang11}.  Discussions on these subjects can be found in \cite{Arons03}.

\subsection{Maximum acceleration energy}

The internal, mainly toroidal magnetic field partially threads the magnetar crust, making up a mainly poloidal magnetosphere with surface dipole strengths $B_{\rm d}\sim 10^{15}$~G that are required to account for the observed spin-down rates \cite{Thompson93}. This dipole component decreases as $B(r) = (1/2)B_{\rm d}(R_*/r)^3$ according to the distance from the star's surface $r$, with $R_*$ the radius of the star. Beyond the light cylinder radius $R_{\rm L}\equiv c/\Omega$, the dipole field structure cannot be causally maintained and the field becomes mostly azimuthal, with field lines spiraling outwards and with strength decreasing as $B(r)\sim B(R_{\rm L})(R_{\rm L}/r)$. The out-flowing relativistic plasma at $r>R_{\rm L}$ (the magnetar ``wind")  thus has magnetospheric voltage drops across the magnetic field of magnitude \cite{Blasi00,Arons03}:
\begin{eqnarray}\label{eq:Phi}
\Phi_{\rm wind} &\sim& rB(r)\sim R_{\rm L}B(R_{\rm L}) = \frac{\Omega^2\mu}{c^2} \\
&=& 3\times 10^{22}\mu_{33}\Omega_4^2~{\rm V}\,, 
\end{eqnarray}
where $\Omega_4\equiv\Omega/10^{4}\,{\rm s}^{-1}$ is the angular velocity of the star and $\mu_{33}\equiv \mu/10^{33}\,{\rm cgs}$ its dipole moment with
$ \mu = B_{\rm d}R_*^3/{2} = 10^{33}~\mbox{cgs}\,(B_{\rm d}/2\times10^{15}~\mbox{G})(R_*/10~{\rm km})^3$.

Assuming that particles with charge $q$ experience a fraction $\eta$ of this voltage drop, they should gain the energy  \cite{Blasi00,Arons03}:
\begin{equation}\label{eq:E_Omega}
E(\Omega) = q\eta\Phi_{\rm wind} = q\eta\frac{\Omega^2\mu}{c^2} = 3\times 10^{21} Z \eta_1\Omega_4^2\mu_{33}~{\rm eV}\, ,
\end{equation}
where we define $\eta_1 \equiv \eta/0.1$. For magnetars beginning their lives with millisecond rotation periods, particles in principle can achieve ultrahigh energies of $E>10^{20}$~eV by acceleration in the wind.

\subsection{Injection spectrum}

Inside the light cylinder, the magnetosphere corotates with the star and the ion density corresponds to the maximum current density of particles extracted from the star surface, i.e., the Goldreich-Julian charge density $\rho_{\rm GJ}$ \cite{Goldreich69}. Assuming that this current is entirely tapped in the wind for acceleration, one can write the instantaneous particle injection rate $\dot{N_{\rm i}}$ as a function of $\Omega$ and thus of the particle energy at a given time $E$. The energy spectrum of the particles accelerated by a magnetar during its spin-down reads:
\begin{equation}\label{eq:spectrum_basic}
\frac{{\rm d} N_{\rm i}}{{\rm d} E} =  \dot{N_{\rm i}}\, \left(-\frac{{\rm d} t}{{\rm d} \Omega}\right) \, \frac{{\rm d} \Omega}{{\rm d} E}\, .
\end{equation}
The spin-down of a magnetar is driven by electromagnetic energy losses and gravitational wave losses (see for example Section~10.5 of Shapiro \& Teukolsky \cite{Shapiro83}). Expressing these losses in terms of $\Omega$ and thus $E$, Eq.~(\ref{eq:spectrum_basic}) can be transformed into \cite{Arons03}:
\begin{equation}\label{eq:spectrum_arons}
\frac{{\rm d} N_{\rm i}}{{\rm d} E} = \frac{9}{4}\frac{c^2I}{Ze\mu E}\left( 1+\frac{E}{E_{\rm g}}\right)^{-1}\, ,
\end{equation}
where the critical gravitational energy at which gravity wave and electromagnetic losses are equal reads:
\begin{equation}\label{eq:Eg}
E_{\rm g} = \frac{5}{72}\frac{Z\eta e\mu^3}{GI^2\varepsilon^2} = 3\times 10^{20} \frac{Z\eta_1\mu_{33}^3}{I_{45}^2\varepsilon_2^2}\,{\rm eV}\, ,
\end{equation}
with $I_{45}\equiv I/10^{45}\,{\rm g\,cm}^2$, the principal moment of inertia of the star. The ellipticity of the magnetar, created by the anisotropic pressure from the interior magnetic field, can be evaluated numerically by: $\varepsilon \sim10^{-2} \, (3B_{\rm d}^2-\langle B_{\rm t}^2\rangle)/(4\times 10^{16}\,{\rm G})$, where the brackets denote a volume average over the entire core (see, e.g., \cite{Bonazzola96,Cutler02}). 

We note $\varepsilon_2 \equiv \varepsilon/10^{-2}$. We neglect the influence of r-mode instabilities on the magnetar spin-down: this effect should mainly affect the cut-off of the injected energy spectrum by modifying the energy loss component due to gravitational wave losses \cite{Arons03}. 

Equation~(\ref{eq:spectrum_arons}) gives the energy spectrum of cosmic rays that are injected by a single magnetar in the interstellar medium, over $1-2$ hours to go down to ankle energies (see Eq.~28 and 29 of Ref.~\cite{Arons03} for evaluations of the spin-down time).

\section{Integrated UHECR flux from magnetar populations}\label{section:integrated}

\begin{figure}[t]
\begin{center}
\includegraphics[width= \columnwidth]{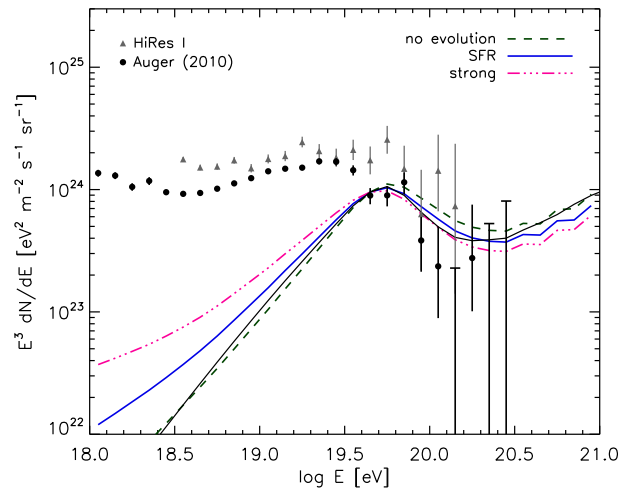} 
\caption{Cosmic ray spectra for an uniform distribution of initial voltages among magnetar parameters (Eq.~\ref{eq:spectrum_uni}). A pure proton composition is injected with the parameters chosen in the numerical application of Eq.~(\ref{eq:Eg}), with initial rotation velocity $\Omega_{\rm i,4}$ (i.e., $E_{\rm max}=3\times 10^{21}$~eV and $E_{\rm g}=3\times 10^{20}$~eV). The flux calculated analytically for no source evolution history (black solid) is compared to the numerical propagation computations with different source evolution models. See text for the magnetar rate $n_{\rm m}$ required to fit the data in each case. The spectra observed by HiRes \cite{Abbasi08} and the Pierre Auger Observatory \cite{Abraham10} are overlaid.}\label{fig:crspec_uni}
\end{center}
\end{figure}

\subsection{Uniform distribution}\label{subsection:uniform}

We calculate the flux of UHECR on Earth produced by a population of magnetars. We first assume that all magnetars have identical physical properties and that they occur at a rate $n_{\rm m}$ per unit volume per year in our local Universe. At energies above the ankle, the influence of the intergalactic magnetic fields on the shape of the UHECR spectrum should be negligible. The cosmic ray spectrum ${\rm d}N/{\rm d}E$ then follows the relation:
\begin{equation}\label{eq:nE}
\frac{\partial}{\partial E}\left(\dot{E}\,\frac{{\rm d}N}{{\rm d}E}\right) = W_{\rm geom}\, n_{\rm m} \,\frac{{\rm d}N_i}{{\rm d} E} \, .
\end{equation}
Here $W_{\rm geom}$ is a geometrical factor that accounts for the fact that all magnetars cannot inject ions from the stars' atmospheres into the wind in the rotational equator (see section~4.4 of Ref.~\cite{Arons03}). For numerical applications, we will choose in the rest of this paper the value $W_{\rm geom}=0.5$. Integrating equation~(\ref{eq:nE}) using $\dot{E} = -E/T_{\rm loss}$ (energy losses due cosmological expansion and to interactions with the cosmological photon backgrounds) and Eq.~(\ref{eq:spectrum_arons}), we get \cite{Arons03}:
\begin{eqnarray}
J(E)&=& \frac{c}{4\pi}\frac{{\rm d}N}{{\rm d}E}\\
&=& W_{\rm geom}\frac{9}{16\pi}\frac{Ic^3}{Ze\mu}\, n_{\rm m,0} \,E^{-1} T_{\rm loss}(E)\times\nonumber\\
&&\ln\left[ \frac{E_{\rm max}}{E} \frac{1+(E/E_{\rm g})}{1+(E_{\rm max}/E_{\rm g})}\right]\, .\label{eq:spectrum_uni}
\end{eqnarray}
$E_{\rm max}$ is the maximum acceleration energy corresponding to the initial rotation velocity: $E_{\rm max}\equiv E(\Omega_{\rm i})$. We assumed in this calculation that the magnetar birth rate $n_{\rm m,0}$ remains constant throughout time. Though magnetars are bursting UHECR sources, the spread in their arrival time on Earth induced by deflections on the intergalactic magnetic fields should be sufficient to account for the continuous detection of particles (see discussion by Ref.~\cite{W95} for the case of gamma-ray bursts, and Section~\ref{section:singlegw} of this paper for estimates of the time delays due to deflections).

Figure~\ref{fig:crspec_uni} presents the UHECR flux calculated analytically following Eq.~(\ref{eq:spectrum_uni}) for magnetars with the parameters chosen in the numerical application in Eq.~(\ref{eq:Eg}), with initial angular velocity $\Omega_{\rm i,4}$ (i.e., $E_{\rm max}=3\times 10^{21}$~eV), and injecting a pure proton composition (black solid line). The calculation of the term $T_{\rm loss}(E)$ due to the cosmological expansion and to energy losses on the cosmological photon backgrounds takes into account photo-pion production and pair production processes on the cosmological microwave background (CMB) and the infrared, optical and ultraviolet background photons modeled by Ref.~\cite{SMS06}.  The rate of magnetars required to fit the observed spectrum is $n_{\rm m, 0}=7\times10^{-8}$~Mpc$^{-3}$~yr$^{-1}$, which corresponds to 3.5\% of the whole magnetar population, if one assumes that a magnetar birth rate of $10^{-4}$ per year per galaxy, and that the average galaxy density is of $2\times10^{-2}$~Mpc$^{-3}$ \cite{Gaensler01}. One can note however that both HiRes \cite{Abbasi09} and the Pierre Auger Observatory \cite{Abraham10} report systematic uncertainties of order 20\% on the absolute energy scale of the spectrum, which should be considered for the evaluation of $n_{\rm m,0}$.

We also present the spectra calculated numerically using the cosmic ray propagation code developed in \cite{Allard05,Allard06,KAM09}, assuming that each source injects particles following Eq.~(\ref{eq:spectrum_arons}), with the same magnetar physical parameters as in the analytical case. Cosmological expansion losses, as well as losses due to interactions of UHECRs with the CMB and the infrared, optical and ultraviolet background photons modeled by Ref.~\cite{SMS06} are included. We examine the effects of three typical source emissivity evolution cases: (i) no evolution, (ii) the source evolution follows the star formation rate normalized to unity at $z=0$ derived in Ref.~\cite{Hopkins06} (labeled SFR):
\begin{equation}
R_{\rm SFR}(z)=\frac{1+7.64z}{1+(z/3.3)^{5.3}}\, ,
\end{equation}
 and (iii) a strong source evolution case that for example Faranoff-Riley type II galaxies might follow \cite{Wall05}. These evolutions are described in detail in Ref.~\cite{KAO10}. The uniform case is, as expected, in good agreement with the analytical calculation. The spectra are fitted by eye to the data and the magnetar birth rate at $z=0$ required in each case are: $n_{\rm m, SFR} \sim 0.8\, n_{\rm m, 0} \sim 5.6\times 10^{-8}$~Mpc$^{-3}$~yr$^{-1}$ for the star formation type evolution, and $n_{\rm m, strong} \sim 0.7\, n_{\rm m, 0} \sim 5\times 10^{-8}$~Mpc$^{-3}$~yr$^{-1}$ for the strong evolution. 

The spectra calculated for an uniform distribution of initial voltage drops among magnetars only mildly fits the observed UHECR spectrum. One would need another type of source to cover the energy range $10^{18-19.5}$~eV, a scenario which is not very attractive in terms of simplicity and fine tuning.  We propose in the next sections to introduce a distribution of the initial voltage drops among magnetars to reconcile the observed spectrum with the magnetar scenario.

\subsection{Distribution of $\Omega_{\rm i}$}\label{subsection:Omegai}

Equation~(\ref{eq:spectrum_uni}) assumes that all magnetars have the same initial voltage $\Phi_{\rm i}=\Omega_{\rm i}^2\mu/c^2$. It is likely however that this voltage differs from star to star, because of different initial angular velocities $\Omega_{\rm i}$, and/or differences in the magnitude of the surface dipole fields and hence on the dipole moment $\mu$. The distribution of both these quantities among the magnetar population is basically unknown. 
One might then relax the previous assumption and introduce a distribution of magnetar birth rates according to the starting voltage, assuming a power-law:
\begin{equation}\label{eq:nm_dPhi}
\frac{{\rm d} n_{\rm m}}{{\rm d}\Phi_{\rm i}} = \frac{n_{\rm m}}{\Phi_{\rm i,max}}\frac{s-1}{(\Phi_{\rm i,max}/\Phi_{\rm i,min})^{s-1}-1}\left(\frac{\Phi_{\rm i}}{\Phi_{\rm i,max}}\right)^{-s}\, ,
\end{equation}
with $\Phi_{\rm i,min}\le \Phi_{\rm i}\le\Phi_{\rm i,max}$.
As a function of the initial acceleration energy $E_{\rm i}$, we get:
\begin{equation}\label{eq:nm_dE}
\frac{{\rm d} n_{\rm m}}{{\rm d}E_{\rm i}} = \frac{{\rm d} n_{\rm m}}{{\rm d}\Phi_{\rm i}} \frac{{\rm d} \Phi_{\rm i}}{{\rm d}E_{\rm i}}  = n_{\rm m}\chi \left(\frac{E_{\rm i}}{E_{\rm i,max}}\right)^{-s}\, ,
\end{equation}
where we defined:
\begin{equation}
\chi\equiv\frac{1}{E_{\rm i,max}}\frac{s-1}{(E_{\rm i,max}/E_{\rm i,min})^{s-1}-1}\, .
\end{equation}

Equation~(\ref{eq:spectrum_uni}) is then transformed into:
\begin{eqnarray}
J(E) &=& \int_{E_{\rm i,min}}^{E_{\rm i,max}}\frac{\partial J(E,E_{\rm i})}{\partial E_{\rm i}}\,{\rm d}E_{\rm i}\ .
\end{eqnarray}
This integral leads to different results according to which parameter $\Omega_{\rm i}$, $\mu$, or both is distributed among the magnetar population. For a distribution of initial angular velocities $\Omega_{\rm i}$ with a fixed dipole moment $\mu$ among the stars, the integral is calculated in Appendix~\ref{appendix:Omegai}. One can note that the slope of the overall UHECR spectrum below the high energy cut-off happens to be $s$, the slope of the distribution of $\Omega_{\rm i}$ among magnetars. 

\begin{figure}[!t]
\begin{center}
\includegraphics[width=\columnwidth]{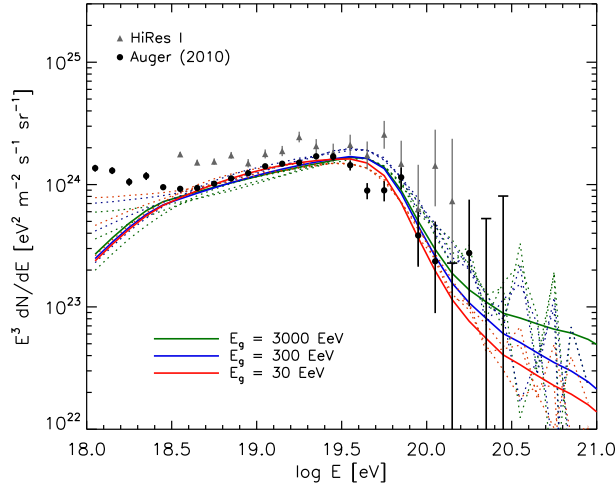} 
\caption{Cosmic ray spectra when distributions of $\Omega_{\rm i}$ are included, for a pure proton injection, $E_{\rm i,min}=3\times10^{18}~$eV, $E_{\rm i,max}=3\times 10^{21}~$eV, $\mu_{33}$, $I_{45}$ and $\eta_1$. Colors indicate $E_{\rm g}=30$, 300 and 3000~EeV respectively. Analytical results from Appendix~\ref{appendix:Omegai} are plotted in solid lines. The results of our numerical particle propagation are shown in dotted lines, for no, SFR and strong evolution models (from bottom to top in the low energy end). The magnetar population parameters used to best fit the observed data are presented in Table~\ref{table:Omegai}.}\label{fig:crspec_Omegai}
\end{center}
\end{figure}

The resulting cosmic ray spectra when distributions of $\Omega_{\rm i}$ are included, for a pure proton injection, $E_{\rm i,min}=3\times10^{18}~$eV, $E_{\rm i,max}=3\times 10^{21}~$eV and $E_{\rm g}=30$, 300 and 3000~EeV are presented in Fig.~\ref{fig:crspec_Omegai}. The analytical results from Appendix~\ref{appendix:Omegai} are plotted in solid lines. The results of our numerical particle propagation are shown in dotted lines, with colors corresponding to the critical gravitational energy $E_{\rm g}$ labeled for the analytical cases. For each $E_{\rm g}$, the spectrum obtained for no evolution, SFR and strong evolution models are shown (from bottom to top in the low energy end). The numerical calculation is performed by assuming that all magnetars inject particles according to Eq.~(\ref{eq:spectrum_arons}), up to a maximum energy that is sampled according to Eq.~(\ref{eq:nm_dE}). The agreement between the simulated and the analytical spectra in the absence of source evolution is good. 

Note that in principle, the value of $E_{\rm g}$ could also depend on the angular velocity $\Omega_{\rm i}$ through the ellipticity $\varepsilon$, as we will see in the next section. This dependence is not trivial as we will discuss, and we assume for simplicity in this section that $E_{\rm g}$ is fixed for all magnetars. 

Table~\ref{table:Omegai} recaps the values of the magnetar population parameters required to fit the observed UHECR spectrum for the various scenarios represented in Fig.~\ref{fig:crspec_Omegai}. The required magnetar birthrate is fairly high, ranging from $10-30$\% of the total supernova rate (of order $10^{-4}$~Mpc$^{-3}$~yr$^{-1}$). The rate of magnetars is widely uncertain, but such values are advocated by some authors~\cite{Gill07}. One may note furthermore that these rates includes mostly neutron stars that are are born with relatively low rotation periods ($\Omega_{\rm i,min}\sim 300$~s$^{-1}$, for $\mu_{33}$). The required rate is slightly lower when a source evolution is included. The values indicated in Table~\ref{table:Omegai} are approximative, as the fits are done by hand and because of the intrinsic uncertainties in the observed data. Varying $E_{\rm i,max}$ has little impact on $n_{\rm m,0}$ and $s$ because most of the population contributes at low energies due to the decreasing power-law distribution.

Finally, Figure~\ref{fig:crspec_Omegai} demonstrates that the three values of $E_{\rm g}$ represented are allowed by the observed data. Low and strong gravitational wave emission models are equally possible at the present experimental stage. Stronger statistics in the highest energy end of the spectrum would help constrain the shape of the cut-off. 

\begin{table}[!bh]
\renewcommand{\arraystretch}{1.2}
\vspace{0.1cm}
\begin{ruledtabular}
  \begin{tabular}{lccc}
 Source evolution 	& $E_{\rm g}$ 	& $n_{\rm m,0}$ 	& $s$\\
 model			&  [EeV] 		& [Mpc$^{-3}$~yr$^{-1}$]& \\[0.1cm]
 \hline
  				&	30			& $2.6\times 10^{-5}$ & 2.2	\\
  no evolution 		& 	300			& $2.7\times 10^{-5}$ & 2.5	\\
  				&	3000			& $2.9\times 10^{-5}$ & 2.6	\\
 \hline
  				&	30			& $1.4\times 10^{-5}$ & 2.0	\\
  SFR	 		& 	300			& $1.4\times 10^{-5}$ & 2.3	\\
  				&	3000			& $1.6\times 10^{-5}$ & 2.4	\\
 \hline
  				&	30			& $7.6\times 10^{-6}$ & 1.8	\\
  strong			& 	300			& $7.6\times 10^{-6}$ & 2.1	\\
  				&	3000			& $8.1\times 10^{-6}$ & 2.2 \\
\end{tabular}
\end{ruledtabular}
\caption{{\bf Parameters for distribution of $\Omega_{\rm i}$} (Section~\ref{subsection:Omegai}). Magnetar density at $z=0$, $n_{\rm m,0}$, and spectral indices $s$ of the distribution of initial voltage drops required to fit the observed UHECR spectrum are indicated for various source model evolutions and critical gravitational energy $E_{\rm g}$. We assume a pure proton injection, $E_{\rm i,min}=3\times10^{18}~$eV, $E_{\rm i,max}=3\times 10^{21}~$eV, $\mu_{33}$, $I_{45}$, and $\eta_1$.}\label{table:Omegai}      
\end{table}

\begin{figure}[!t]
\begin{center}
\includegraphics[width=\columnwidth]{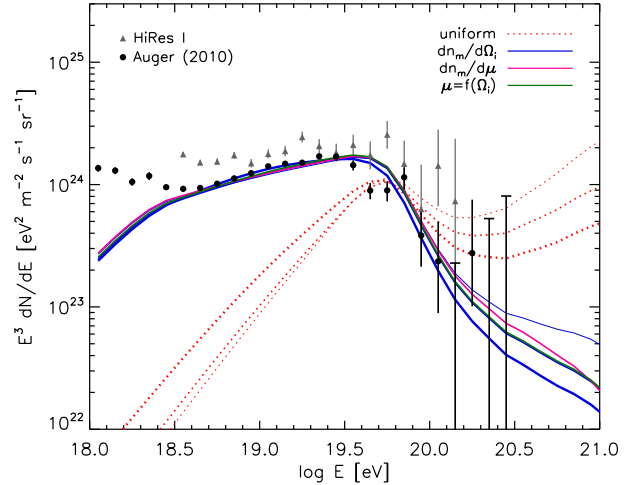} 
\caption{Cosmic ray spectra calculated analytically when distributions of $\Omega_{\rm i}$, $\mu$ and $\mu=f(\Omega_{\rm i})$ are included (different colors), for a pure proton injection, $E_{\rm i,min}=3\times10^{18}~$eV, $E_{\rm i,max}=3\times 10^{21}~$eV, $\mu_{33}$, $I_{45}$ and $\eta_1$. For the distribution of $\Omega_{\rm i}$, cases for $E_{\rm g}=30$, 300 and 3000~EeV are represented with decreasing thickness. The magnetar population parameters used to best fit the observed data are presented in Table~\ref{table:manydistr}.}\label{fig:crspec_manydistr}
\end{center}
\end{figure}

\subsection{Distribution of $\mu$}\label{subsection:mu}

We now assume that $\Omega_{\rm i}$ is fixed and that the dipole moment $\mu$ has a distribution among magnetars that lead to the distribution in voltages described by Eq.~(\ref{eq:nm_dPhi}). This case is interesting, as the injection spectrum of single magnetars themselves scales inversely to $\mu$ (Eq.~\ref{eq:spectrum_arons}). A weaker $\mu$ will thus lead naturally to stronger fluxes and varying $\mu$ over $[\mu_{\rm i,min},\mu_{\rm i,max}]$ softens the integrated cosmic ray spectrum without requiring higher source densities to account for the low energy flux. 

The critical gravitational energy $E_{\rm g}$ also depends on $\mu$ (Eq.~\ref{eq:Eg}). However, this dependency is not straightforward as the ellipticity $\varepsilon$ is also expected to depend on the magnetic field. For magnetic field induced deformation of the star, assuming that the star interior is a compressible perfect fluid, one can consider that the ellipticity can be written \cite{Bonazzola96}:
\begin{equation}\label{eq:eps}
\varepsilon = \beta\frac{R_*^2}{GI^2}\mu^2\ .
\end{equation}
Here, $\beta$ is the magnetic distortion factor introduced by Ref.~\cite{Bonazzola96}, which measures the efficiency of the interior magnetic field in distorting the star. This factor depends on the equation of state of the star interior and on its magnetic field geometry. Ref.~\cite{Bonazzola96} find that the value of $\beta$ can range between $1-10$ for perfectly conducting interiors (normal matter), $10-100$ for type I superconductors and can reach $\gtrsim 100$ for type II superconductors. 
The relation above might not stand however when the interior magnetic field is dominated by the toroidal component \cite{Cutler02}, and the magnetar internal field being already close to or exceeding the critical field value (of order $10^{15}$\,G \cite{Easson79}), it is possible that no significant enhancement of the magnetar deformation happens beyond $\varepsilon\sim 10^{-4}-10^{-3}$.

In the following, we will rely on this relation for simplicity. The influence of $E_{\rm g}$ (hence of the dependency between $\varepsilon$ and $\mu$) is only noticeable at the very high energy end of the UHECR spectrum ($E\gtrsim$ few $10^{20}$~eV), with variations that could not be measured with the observational statistics of current detectors. Differences in $E_{\rm g}$ have an impact on the normalization $n_{\rm m}$ that are below the systematic uncertainties of the observed UHECR spectrum. 

The analytical spectrum for a distribution of $\mu$ among magnetars is calculated in Appendix~\ref{appendix:mu}. In this case, the spectral index of the UHECR spectrum before the high-energy cut-off is given by $s+1$.
The results are presented in Fig.~\ref{fig:crspec_manydistr} for a pure proton injection, no source evolution, $\Omega_{\rm i}=10^4$~s$^{-1}$, $E_{\rm i,min}=3\times10^{18}~$eV, $E_{\rm i,max}=3\times 10^{21}~$eV and $\beta=700$, which corresponds to $\varepsilon\sim 10^{-2}$. The corresponding values for $n_{\rm m,0}$ and $s$ are presented in the third row of Table~\ref{table:manydistr}. The shape of the spectra is similar to that obtained for a distribution of $\Omega_{\rm i}$ and superimpose well in the low energy region. This is explained by the fact that the spectra calculated in Appendix~\ref{appendix:Omegai} and \ref{appendix:mu} have nearly the same expression, the main difference residing in the normalization. With lower values of $\beta$, one recovers the shape of the spectra calculated for high $E_{\rm g}$ in the previous section. In Table~\ref{table:manydistr}, we have also indicated the normalization and spectral index values required to fit the observed spectrum for $\beta=200$, 700 and 2000. These values correspond to $E_{\rm g}\sim 3000$, 300 and 30~EeV respectively, for $I_{45}$ and $\mu_{33}$. One can notice that the influence of $\beta$ on $n_{\rm m,0}$ and $s$ is very mild. 

Here, the magnetar birth rates (in Table~\ref{table:manydistr}) needed to fit the observed UHECR spectrum are lower of two orders of magnitude compared to the case treated in the previous section. The rates are close to those calculated for the uniform distribution. This case is less tight in terms of population density as compared to the distribution in rotation velocities, but necessitates that all the magnetars considered are sub-millisecond rotators at birth. It is also ``disappointing" in terms of gravitational wave signatures, as it should lead to faint signals, as we will see in Section~\ref{subsection:gwmu}.

\begin{table}[th]
\renewcommand{\arraystretch}{1.2}
\vspace{0.1cm}
\begin{ruledtabular}
  \begin{tabular}{lccccc}
 Distribution 	& $E_{\rm g}$& $\beta$	& $\alpha$ &$n_{\rm m,0}$ 	& $s$\\
 model			&  [EeV] 	&	& [G s] & [Mpc$^{-3}$~yr$^{-1}$]& \\[0.1cm]
  \hline
  										&	30		&-	&-& $3\times 10^{-8}$ & -	\\
  uniform ($\Omega_{\rm i,4}$, $\mu_{33}$)		& 	300		&-	&-& $7\times 10^{-8}$ & -	\\
  										&	3000		&-	&-& $4\times 10^{-8}$ & -	\\
 \hline
  												&	30		&-	&-& $3\times 10^{-5}$ & 2.2	\\
  ${\rm d}n_{\rm m}/{\rm d}\Omega_{\rm i}$ ($\mu_{33}$)		& 	300		&-	&-& $3\times 10^{-5}$ & 2.5	\\
  												&	3000		&-	&-& $3\times 10^{-5}$ & 2.6	\\
 \hline
 												& 	-	&200			&-& $8\times 10^{-8}$	&1.6\\
 ${\rm d}n_{\rm m}/{\rm d}\mu$	($\Omega_{\rm i,4}$)		& 	-	&700		&-& $8\times 10^{-8}$ & 1.5	\\
  												& 	-	&2000		&-& $8\times 10^{-8}$	&1.5\\

 \hline
  $\mu=f(\Omega_{\rm i})$			& 	300	&-	&$10^{11}$	& $5\times 10^{-7}$ & 2.2	\\
  												& 	300	&-	&$10^{13}$	& $1\times 10^{-5}$ & 2.2\\
\end{tabular}
\end{ruledtabular}
\caption{{\bf Parameters for different distributions}. Magnetar density at $z=0$, $n_{\rm m,0}$, and spectral indices $s$ of the distribution of initial voltage drops required to fit the observed UHECR spectrum in absence of source evolution are indicated for different distribution models, critical gravitational energy $E_{\rm g}$, distortion parameter $\beta$ and $\alpha$ as defined in Eq.~(\ref{eq:B_Omega}). We assume a pure proton injection, $E_{\rm i,min}=3\times10^{18}~$eV, $E_{\rm i,max}=3\times 10^{21}~$eV, $I_{45}$, and $\eta_1$.}\label{table:manydistr}
\end{table}

\subsection{Distribution of $\mu=f(\Omega_{\rm i})$}\label{subsection:muf}

It is plausible that the surface dipole magnetic field $B_{\rm d}$ and its initial angular velocity $\Omega_{\rm i}$ are not independent. Ref.~\cite{Duncan92} conjectures that an efficient $\alpha\omega$-dynamo operates during the formation of the magnetar. For proto-neutron stars born with initial periods $P_{\rm i}$, one can calculate that the saturation magnetic field (when energy equipartition is reached between the fluid and the field) on small scales, generated by kinematic growth of seed fields under dynamo action, is of order $B_{\rm sat}\sim3\times10^{17}~{\rm G}\,(1~{\rm ms}/P_{\rm i})$. The observed large scale dipole field $B_{\rm d}$ can be regenerated from this field differential rotation and convection processes. One can thus assume $B_{\rm d}$ to be a constant fraction of the saturation field and write the following relation~\cite{Xu02}:
\begin{equation}\label{eq:B_Omega}
B_{\rm d}=\alpha\frac{\Omega_{\rm i}}{\pi}\, ,
\end{equation}
where we will choose arbitrarily $\alpha \in[10^{11},10^{13}]~\mbox{G}\,\mbox{s}$. The maximum value of $\alpha$ is chosen to account for the order of magnitude of the strongest dipole fields suggested to exist in magnetars, assuming $\Omega_{\rm i}=10^{4}$~s$^{-1}$. 

In reality, it is likely that the relation above be modified, for example if the ratio between $B_{\rm d}$ and $B_{\rm sat}$ is a function of $P_{\rm i}$ (see the corresponding discussion in Ref.~\cite{Xu02}). Because these dependencies are not elucidated, and for the sake of simplicity, we will stick for the time being to the above formula. 

The situation is then similar to the previous section: magnetars birthrates are distributed according to $\mu$, with a different relation between $\Phi_{\rm i}$ (or $E_{\rm i}$) and $\mu$. The angular velocity as a function of $\mu$ reads:
\begin{equation}
\Omega_{\rm i} = \frac{2\pi}{\alpha R_*^3}\mu\ .
\end{equation}
The initial voltage can then be expressed as:
\begin{equation}
\Phi_{\rm i} = \frac{A}{q\eta} \mu^{3} \quad \mbox{with}\quad A\equiv{q\eta}\frac{4\pi^{2}}{\alpha^2 R_*^6 c^2}\, .\label{eq:Phialpha}
\end{equation}
The analytical calculation of the spectrum is then straightforward and can be found in Appendix~\ref{appendix:muOmegai}.
We assume in this calculation that $E_{\rm g}$ is fixed among magnetars to avoid a complicated cut-off function at the highest energies that would not affect strikingly the normalization of the spectrum, nor the spectral index of the distribution of initial voltages as discussed in Section~\ref{subsection:mu}. Also because, as discussed in the same section, the dependency of $\varepsilon$ on $\mu$ is not clearly determined. 

As expected, the spectrum found matches perfectly the one calculated for a distribution of $\Omega_{\rm i}$ (see Fig.~\ref{fig:crspec_manydistr}).
The value of $n_{\rm m,0}$ scales as $\alpha^{-2/3}$.

In Table~\ref{table:manydistr}, we calculated the values of the normalization and the spectral index needed to fit the observed spectrum in absence of source evolution. The cases for SFR and strong source evolutions can be roughly extrapolated through the following relations, as we checked in Section~\ref{subsection:uniform} and with supplementary simulations: $n_{\rm m,0, SFR} \sim 0.8\, n_{\rm m, 0, no evol}$, and $s_{\rm SFR}\sim s_{{\rm no evol}}-0.2$,  for the star formation type evolution; $n_{\rm m,0, strong} \sim 0.7\, n_{{\rm m, 0, no evol}} \sim 5\times 10^{-8}$~Mpc$^{-3}$~yr$^{-1}$, and $s_{\rm strong}\sim s_{{\rm no evol}}-0.4$ for the strong evolution.

\section{Implication for the diffuse gravitational wave signal}\label{section:gw}

\subsection{Gravitational waves from a single magnetar}\label{section:singlegw}

Rotating neutron stars are expected to emit copious amounts of gravitational radiation, mainly at its rotation frequency and at twice its rotation frequency, provided that it deviates from axi-symmetry (\cite{Bonazzola94} for a review). In the case of a magnetar, the deviation from axi-symmetry is believed to be caused principally by the neutron star's internal magnetic field \cite{Bonazzola96}. The energy losses from gravitational waves for a star with ellipticity $\varepsilon$ can be estimated as follows \cite{Ostriker69}:
\begin{equation}
\dot{E}_{\rm gw} = \frac{32}{5}\frac{GI^2\varepsilon^2\Omega^6}{c^5}\ .
\end{equation}
The gravitational spectral energy emitted by a single source between frequency $\nu\equiv\Omega/(2\pi)$ and $\nu+{\rm d}\nu$ thus reads:
\begin{eqnarray}
\frac{{\rm d}E_{\rm gw}}{{\rm d}\nu} &=&2\pi\,\frac{\dot{E}_{\rm gw}}{2\dot{\Omega}} \\
&=& K\,\nu^3\left[1+\frac{K}{\pi^2I}\nu^2\right]^{-1} \mbox{with} \quad \nu\in[0,\nu_{\rm i}]\ ,\label{eq:dEgw}
\end{eqnarray}
where $\nu_{\rm i}\equiv \Omega_{\rm i}/\pi$ is twice the initial spin frequency and
\begin{eqnarray}
K&\equiv& \frac{288\pi^4}{5}\frac{GI^3\varepsilon^2}{c^5\mu^2} =\frac{72\pi^4}{5}\frac{\beta^2R_*^4}{c^2GI}\mu^2\, .\label{eq:K}\\
\end{eqnarray}
The last equality assumes that the deformation of the magnetar is due to its internal magnetic structure, and that it is a quadratic function of the amplitude of the magnetic dipole moment, as in Eq.~(\ref{eq:eps}). 

The detectability of gravitational wave signals from individual magnetars has been investigated by several authors \cite{Stella05,Dallosso07} who find that the signals could be detectable by Advanced LIGO-class detectors up to the distance to the Virgo Cluster. Though such a detection would be a watershed in several fields of high-energy astrophysics, note that gravitational radiation from single magnetars cannot be observed in coincidence with UHECR events. 
The delay induced by extragalactic magnetic fields of mean strength $B$ and coherence length $\lambda_B$ on particles of charge $Z$ and energy $E$ with respect to gravitational waves over a distance $D$ reads~\cite{AH78,KL08b}:
\begin{eqnarray}\label{eq:delay}
\delta t\,\simeq\,2.3\times 10^2\,{\rm yrs}\,
Z^2\left(\frac{D}{10\,{\rm Mpc}}\right)^2\,\left(\frac{\lambda_B}{0.1\,{\rm Mpc}}\right)\times\nonumber\\
\,\left(\frac{E}{10^{20}\,{\rm eV}}\right)^{-2}\left(\frac{B}{10^{-9}\,{\rm G}}\right)^2 .
\end{eqnarray}
It is likely that magnetic fields at this level of $\sim\,$nG are present in our local supercluster, over a few megaparsecs. 
For {\it homogeneous} intergalactic magnetic fields of lower overall strength ($B\lesssim 10^{-12}$~G), the time delay could be shorter than a year over 100~Mpc. However, the crossing of one single magnetized filament (size $\bar{r}_i$, field strength $B$ and coherence length $\lambda_i$) will lead to a slight deflection that will induce a time delay with respect to a straight line of order \citep{AH78,WM96,Harari02a,KL08b}: 
\begin{eqnarray}
\delta t_i\,\simeq\,10^4\,{\rm yrs}\,
Z^2\left(\frac{\bar r_i}{2\,{\rm Mpc}}\right)^2\,\left(\frac{\lambda_i}{0.1\,{\rm Mpc}}\right)\times\\
\,\left(\frac{E}{10^{20}\,{\rm eV}}\right)^{-2}\left(\frac{B}{10^{-8}\,{\rm G}}\right)^2 .
\end{eqnarray}
For the same reason, it is unlikely that even for magnetars inside our Galaxy, gravitational wave signals could be detected in coincidence with UHECRs. 

\subsection{Diffuse gravitational wave signals expected for magnetar populations that are sources of the observed UHECR flux}

The possibility of detecting gravitational wave backgrounds generated by an ensemble of neutron stars and of magnetars more specifically, has been discussed by several authors \cite{Giazotto97,Regimbau00,Regimbau01,Regimbau06,Marassi10}. The background results from the superposition of the signals of the population of emitters, integrated over the whole history of their evolution. 

The spectrum of the gravitational stochastic background can be characterized by the quantity $\Omega_{\rm gw}(\nu_0)$, defined as the present-day energy density per logarithmic frequency interval, in gravitational waves of frequency $\nu_0$, divided by the critical energy density of the Universe $\rho c^2$. For a population of sources, each emitting gravitational waves according to Eq.~(\ref{eq:dEgw}), this quantity can be expressed \cite{Allen99,Phinney01}:
\begin{eqnarray}\label{eq:Ogw}
\Omega_{\rm gw}(\nu_0)  &=& 5.7\times 10^{-56} \left(\frac{0.7}{h_0}\right)^2 \,\nu_0   \int_{\nu_{\rm i,min}}^{\nu_{\rm i,max}}  \frac{{\rm d} n_{\rm m}}{{\rm d}\nu_{\rm i}} \,{\rm d}\nu_{\rm i} \times\nonumber\\
&&\hspace{-0.7cm}\, \int_0^{z_{\rm sup}(\nu_{\rm i})} \frac{R_{\rm evol}(z)}{(1+z)^2U(z)}\frac{{\rm d}E_{\rm gw}}{{\rm d}\nu}[\nu_0(1+z)]\,{\rm d}z\, ,
\end{eqnarray}
where $\nu_0$ and $\nu=\nu_0(1+z)$ are the frequencies in the observer and in the source frame respectively, $U(z)\equiv[\Omega_\Lambda+\Omega_{\rm m}(1+z)^3]^{1/2}$, $R_{\rm evol}(z)$ is the dimensionless source evolution rate normalized to 1 at $z=0$. The integral upper bound $z_{\rm sup}(\nu_{\rm i})$ is given by:
\begin{equation}
z_{\rm sup}(\nu_{\rm i})=\left\{
  \begin{array}{ll}
   z_{\rm max} &\quad \mbox{if } \nu_0<{\displaystyle \frac{\nu_{\rm i}}{1+z_{\rm max}}}\\
    {\displaystyle \frac{\nu_{\rm i}}{\nu_0}-1} &\quad\mbox{otherwise,} \\
  \end{array}
\right.
\end{equation}
where  $z_{\rm max}=6$. A distribution of the source density according to the initial frequency $\nu_{\rm i}$ (corresponding to a distribution in $\Phi_{\rm i}$) has been introduced in the calculation of the background signal. 

The number of magnetars is large enough for the time interval between events to be small compared to the duration of a single event. One can show indeed that the duty cycle 
\begin{equation}
\Delta(z)\equiv \int_0^z n_{\rm m,0}R_{\rm evol}(z') \tau_{\rm gw} (1+z')\frac{{\rm d}V}{{\rm d}z}(z')\,{\rm d}z'\ ,
\end{equation} 
defined as the ratio between the  duration of the events and the time interval between successive events satisfies the condition $\Delta\gg 1$. $\tau_{\rm gw}$ is the average duration of a signal produced by a magnetar; the assumption that this duration is of the order of the magnetar spin-down time leads to $\Delta>10^3$. Their gravitational wave signal can thus be considered as a continuous background. Such backgrounds obey the Gaussian statistic and are completely determined by their spectral properties. 

The optimal strategy to detect these signals is to perform a correlation between two or more detectors, possibly widely separated to minimize common noise sources \cite{Christensen92,Flanagan93,Allen99}. The cross-correlation of two ground-based third generation LIGO-type interferometers (LIGOIII or the Einstein Telescope) would lead to sensitivities interesting in the scope of this study \cite{Buonanno03,Regimbau11}. Sensitivities would be further increased with the subsequent generation of experiments such as BBO and DECIGO, which should work in correlation mode by themselves, thanks to multiple detection units \cite{Corbin06,Harry06,Kawamura06}.

\begin{figure}[t]
\begin{center}
\includegraphics[width=\columnwidth]{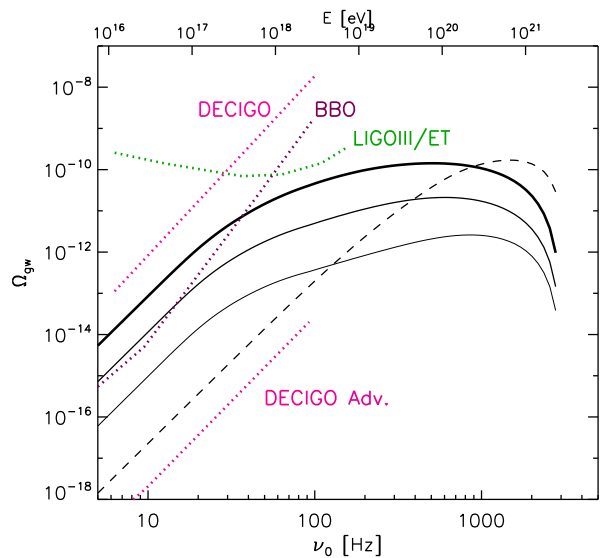} 
\caption{Energy density of the stochastic gravitational wave background $\Omega_{\rm gw}$ produced by magnetars as a function of the observed frequency, assuming a distribution of $\Omega_{\rm i}$ among magnetars, as in Section~\ref{subsection:Omegai}. Solid black lines with increasing thickness: $E_{\rm g}=3000, 300, 30$~EeV, and $n_{\rm 0,m}=0.8\,n_{\rm m,0, noevol}$, and $s=s_{\rm noevol}-0.2$, with $n_{\rm m,0, noevol}$ and $s_{\rm noevol}$ as indicated in the second row of Table~\ref{table:manydistr}. Black dashed line: standard spectrum (Eq.~\ref{eq:Ogw}) with uniform distribution of initial voltages among magnetars, $E_{\rm g}=300$~EeV, $\Omega_{\rm i,4}$ and $n_{\rm m,0}=5.6\times 10^{-8}$~Mpc$^{-3}$~yr$^{-1}$. All cases are for a pure proton injection, $E_{\rm i,min}=3\times10^{18}~$eV, $E_{\rm i,max}=3\times 10^{21}~$eV, $\mu_{33}$, $I_{45}$, and $\eta_1$.
The green dotted line represents the LIGOIII and the Einstein Telescope (ET) approximate sensitivities \cite{Buonanno03,Marassi10}, the pink dotted lines the DECIGO and DECIGO Advanced sensitivities \cite{Kawamura06} and the purple dotted line the BBO sensitivity \cite{Corbin06,Harry06}, all in correlation modes. The values on the upper $x$ axis represent the energy $E=3\times 10^{21}Z\eta_1\mu_{33}(\pi\nu/10^4\,{\rm s}^{-1})^2$~eV. }\label{fig:gw_spec}
\end{center}
\end{figure}

In what follows, we discuss the gravitational wave background spectra obtained for each distribution of magnetars discussed in Section~\ref{section:integrated}, assuming that they are the sources of the observed UHECR flux. The analytical calculations of these spectra are presented in Appendix~\ref{appendix:gw}. The integrals are computed numerically.

We present the cases of an SFR-type source evolution, which can be intuitively expected for a magnetar evolution history, as magnetars are believed to form principally in star-forming regions \cite{Woods06}. We conjecture that the parameters indicated in Table~\ref{table:manydistr} in absence of evolution can be adapted for the SFR case by applying the rules stated at the end of Section~\ref{subsection:muf}.

\begin{figure}[t]
\begin{center}
\includegraphics[width=\columnwidth]{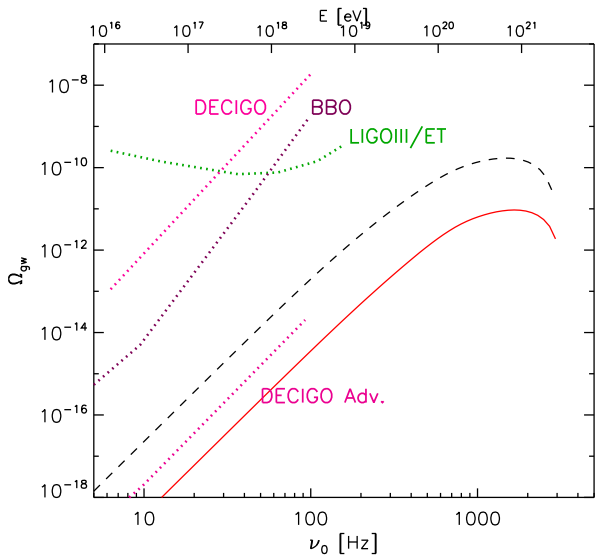} 
\caption{Red solid line: energy density of the stochastic gravitational wave background $\Omega_{\rm gw}$ produced by magnetars as a function of the observed frequency, assuming a distribution of $\mu$ among magnetars and a fixed $\nu_{\rm i}=\nu_{\rm max}=10^4/\pi$, as in Section~\ref{subsection:mu}. We assume a pure proton injection, $\beta=700$, $E_{\rm i,min}=3\times10^{18}~$eV, $E_{\rm i,max}=3\times 10^{21}~$eV, $I_{45}$, and $\eta_1$. We take $n_{\rm 0,m}=0.8\,n_{\rm m,0, noevol}$, and $s=s_{\rm noevol}-0.2$, with $n_{\rm m,0, noevol}$ and $s_{\rm noevol}$ as indicated in the third row of Table~\ref{table:manydistr}. Black dashed line as in Fig.~\ref{fig:gw_spec}. }\label{fig:gw_spec_B}
\end{center}
\end{figure}

\begin{figure}[t]
\begin{center}
\includegraphics[width=\columnwidth]{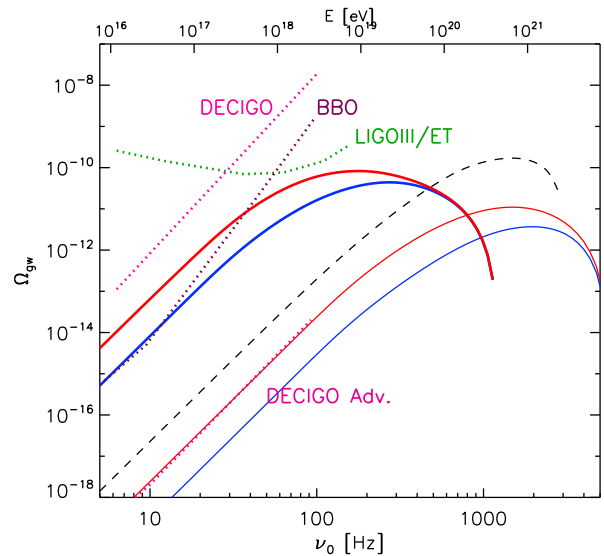} 
\caption{Energy density of the stochastic gravitational wave background produced by magnetars as a function of the observed frequency, assuming $B_{\rm d}=\alpha\nu_{\rm i}$ (Eq.~\ref{eq:B_Omega}), with $\alpha=10^{11}$ and $10^{13}$~G~s for the thin and thick lines respectively, and for $\beta=700$ and 2000 for blue and red lines respectively. We assume a pure proton injection, $E_{\rm i,min}=3\times10^{18}~$eV, $E_{\rm i,max}=3\times 10^{21}~$eV, $I_{45}$, and $\eta_1$. The parameters $n_{\rm 0,m}$ and $s$ are as indicated in the fourth row of Table~\ref{table:manydistr} with the adequate transformations to obtain the SFR evolution case. Black dashed line as in Fig.~\ref{fig:gw_spec}. }\label{fig:gw_spec_real}
\end{center}
\end{figure}

In Figures~\ref{fig:gw_spec}, \ref{fig:gw_spec_B} and \ref{fig:gw_spec_real}, the black dashed lines represent the gravitational wave spectrum obtained for the standard magnetar population scenario, for which the initial voltages are uniformly distributed. It corresponds to the signal computed by Refs.~\cite{Regimbau06,Marassi10}, though the normalization is different. As it appears in Eq.~(\ref{eq:Ogw}), the quantity $\Omega_{\rm gw}$ scales as $n_{\rm m,0}$, and it is hence straightforward to estimate the values that $\Omega_{\rm gw}$ could take for other source densities. In this study, we work under the assumption that magnetars are emitters of the observed UHECRs, which leads to the present scalings for gravitational waves.

Figure~\ref{fig:gw_spec} presents the gravitational wave spectra $\Omega_{\rm gw}$ obtained for different distributions of $\Omega_{\rm i}$ among magnetars, with $\mu$ fixed. The higher contribution of the low initial angular velocity sources (needed to fit the UHECR spectrum) enhances the gravitational wave spectrum at low frequencies, especially in the range $\nu_0\sim1-100$~Hz that will be probed by future instruments in correlation mode. The poloidal field suggested from the observation of Galactic magnetars is of order $\sim 10^{14}$~G, implying that $\varepsilon\sim 10^{-4}-10^{-3}$ ($E_{\rm g}\gtrsim 3000$) would be conservative values. The thin line of Fig.~\ref{fig:gw_spec} shows that the gravitational wave signal would then still be a couple of orders of magnitude above the standard prediction, though below reach of the future instruments. The magnetar ellipticity is however highly uncertain (see e.g., \cite{Braithwaite09}) and more optimistic values of the critical gravitational energy: $E_{\rm g}=30$ and 300~EeV (corresponding to $\varepsilon\sim 0.3$ and $10^{-2}$ respectively) could be considered. The signal could then be strong enough to be detected by the future BBO and DECIGO, and could be close to detection for third generation LIGO-type instruments (LIGOIII or the Einstein Telescope). 

Note that the fact that a background appears to partially overlap to the sensitivity curve of a given detector does not necessarily imply that the signal is detectable. A calculation of the signal-to-noise ratio assuming a given observation time (as done in Refs.~\cite{Allen99,Marassi10}) is needed to thoroughly evaluate the detectability. Such estimates are beyond the scope of this paper, in particular as the design of the instruments discussed here is still highly uncertain.

In Figure~\ref{fig:gw_spec_B}, we plot the case where a distribution of $\mu$ is taken among magnetars, with a fixed initial angular velocity $\Omega_{\rm i}$, and thus a fixed $\nu_{\rm i}=\nu_{\rm max}$. In this situation, the dipole moment being independent of the gravitational radiation frequency, we expect $\Omega_{\rm gw}$ to behave in the same way as for the uniform distribution case. The red dashed line in Fig.~\ref{fig:gw_spec_B} arbors indeed the same shape as the one calculated in Eq.~(\ref{eq:Ogw}). There is a factor $\sim (\mu_{\rm min}/\mu_{\rm max})^{-s}\mu_{\rm min}(\chi q\eta\pi^2\nu_{\rm max}^2/c^2)(s-1)^{-1}$ of difference with the case of uniform distribution of voltages among magnetars (see Eq.~\ref{eq:int_mu}). This factor is smaller than unity for our choices of parameters. This case is therefore ``disapointing" in terms of gravitational wave signal, as it is expected to produce a signal weaker than in the standard magnetar population scenario. One may albeit point out that a sheer distribution of surface dipole field strengths $B_{\rm d}$, uncorrelated with the angular velocities, might not be favored in the current magnetar formation models. 

Finally, Figure~\ref{fig:gw_spec_real} shows an application for distributions of initial voltages among magnetars, with a dependence between the dipole magnetic field strength of the star and its initial angular velocity [$\mu=f(\Omega_{\rm i})$]. We adopted Eq.~(\ref{eq:B_Omega}) as an example of such a relation, and computed the background gravitational wave spectrum for two values of $\alpha$. High values of $\alpha$ imply that high source densities are necessary to fit the UHECR spectrum and enhance consequently the gravitational wave signal. For the most optimistic cases, the signal lies above BBO and DECIGO sensitivities, and is close to the sensitivity of third generation LIGO-type instruments. For low values of $\alpha$ however, the signal drops below the one emitted by a standard population of magnetars. The ellipticities represented here are also optimistic ($\varepsilon\sim 10^{-2}$ for blue lines and $0.3$ for red lines). More reasonable deformations of order $\varepsilon\sim10^{-3}$ would lead to a signal lower of about one order of magnitude.

\section{Discussion and conclusion}\label{section:discussion}

We have presented results for UHECR spectra produced by newly born magnetars and their associated gravitational wave signatures assuming pure proton injection. The metal-rich supernova environment in which the magnetars form supports however the idea that a mixed or iron-rich cosmic-ray composition might be injected. As argued by Refs.~\cite{Ruderman75,Arons79}, strong electric fields could strip iron nuclei off the surface of the young neutron star. The magnetar magnetosphere could thus be enriched in iron-peak elements that can be accelerated via the unipolar induction mechanisms evoked in Section~\ref{section:energetics}. The escape of such accelerated nuclei is not obvious though: the rough estimates of Ref.~\cite{Blasi00} show that ultrahigh-energy iron nuclei could survive the crossing of a supernova envelope under certain conditions, while Ref.~\cite{Arons03} proposes a different scenario in which it is suggested that only protons could escape. Given the recent results of Auger that could be interpreted as an indication of a heavy composition at the highest energy end of the spectrum \cite{Abraham:2010yv}, it will be of interest to investigate further the case of heavy nuclei. 

If a composition different from pure proton were injected, the spectral index necessary to fit the observed UHECR spectrum would be globally harder, implying that the distribution index of the potential drop among magnetars $s$ should be changed accordingly (on average, one would need $s'\sim s-0.2$ in each case). This discussion is valid also when different source evolutions are assumed: this should lead to a change in $s$, thus in a slight change in the normalization of the overall gravitational wave signal. The gravitational wave signature should then be slightly increased. On the other hand, the critical gravitational energy $E_{\rm g}$ scales as the particle charge $Z$: this should lead to a weaker gravitational wave production, all other parameters being fixed. Overall, the signal is not expected to be drastically modified in case of heavy nuclei injection. \\

The single source injection spectrum calculated in Eq.~\ref{eq:spectrum_arons} assumes a mono-energetic acceleration at a given time. Though this assumption is roughly valid for a pure electrostatic acceleration in the magnetar gap region, it might no longer stand for acceleration in the magnetar wind zone. More complex mechanisms than the toy induction model mentioned in Section~\ref{section:energetics} should be at play. Wake-field acceleration or other stochastic acceleration mechanisms as suggested by \cite{Chen02,Arons03,Kuramitsu08,Murase09} could lead to a natural $E^{-2}$ spectrum. Other models such as magnetic reconnection models predict slopes in $-1$. The initial injection spectrum is at the moment uncertain and a thorough study of the acceleration mechanism that might be happening in the wind will be necessary. This paper demonstrates that even extreme injection cases can be reconciled with a proper distribution of parameters among the magnetar population.\\

\begin{figure}[t]
\begin{center}
\includegraphics[width=\columnwidth]{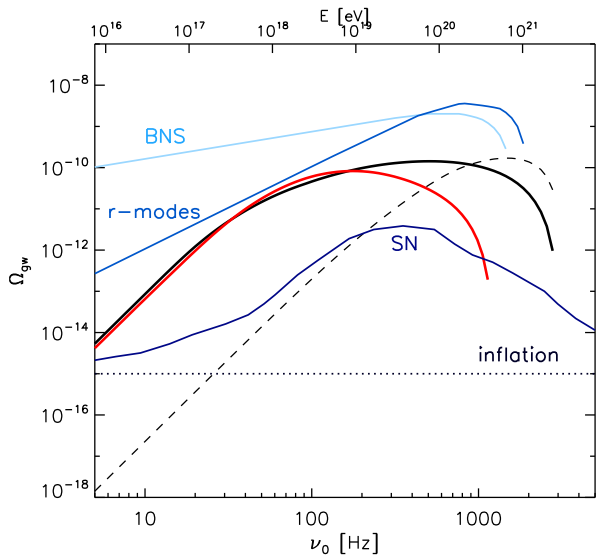} 
\caption{Comparison with energy densities of the gravitational wave backgrounds produced by other astrophysical or cosmological mechanisms. Black solid line: magnetar population with distribution of initial voltages assuming $B_{\rm d}=\alpha\nu_{\rm i}$, with $\alpha=10^{13}$, red solid line: magnetar population with distribution of initial angular velocities $\Omega_{\rm i}$ with $\mu$ fixed, for $E_{\rm g}=30$~EeV, all the other parameters being as in Figs.~\ref{fig:gw_spec} and \ref{fig:gw_spec_real}, black dashed line: standard uniform case as in Fig.~\ref{fig:gw_spec}. Backgrounds from binary neutron star coalescence (BNS) from \cite{Regimbau08},  r-modes 
assuming that 1\% of newly born neutron stars cross the instability window from \cite{Regimbau11}, core-collapse supernov\ae{} from \cite{Buonanno05}.  Horizontal dotted line: maximum version of the gravitational wave stochastic spectrum produced during slow-roll inflation assuming a ratio of the tensorial to the scalar contributions to the cosmic microwave background radiation anisotropy $T/S = 0.3$ and $\pm10^{-3}$ for the running of the tensorial power-law index \cite{Turner97}.}\label{fig:gw_bg}
\end{center}
\end{figure}

Figure~\ref{fig:gw_bg} compares the gravitational wave backgrounds obtained in the present paper with the ones produced by other astrophysical or cosmological sources. A review on these backgrounds can be found in Ref.~\cite{Regimbau11}. Unless the rates of binary neutron star coalescence in the Universe is over-estimated by many orders of magnitude, that signal should lie above the expected magnetar signals. The signal expected from r-mode instabilities in young fast-rotating neutron stars is less established, but under the optimistic assumptions taken in Fig.~\ref{fig:gw_bg}, it could also lie above the magnetar spectrum. Most of the other astrophysical sources, including core-collapse supernov\ae, should produce weaker signals than the optimistic (though reasonable) magnetar populations invoked in this study. A good knowledge of the strength, frequency 
range, and statistical properties of overlapping backgrounds could help distinguish them from the signals that lie beneath. \\

With improved statistics at the highest energies, and an increased anisotropy signal, one might be able to establish if the sources of UHECRs are continuously emitting, or transient objects (see, e.g., \cite{Kalli11}). In the latter case, long gamma-ray bursts (GRBs) and the birth of magnetars would be the best remaining candidates. The distinction between these two sources from their spatial distribution in the sky would be difficult: as transient sources, they should not be visible in the arrival direction of ultrahigh-energy events, and as they form in similar environments, they should present close spatial distributions. 

The detection of specific gravitational wave signatures might be one way of making this distinction. Long gamma-ray bursts are not believed to be strong gravitational-wave emitters, as the signal due to a burst (i.e., by collision of particles) should not be collimated and be too dilute to be observed by any instrument (e.g., \cite{Piran04}). It is possible that the progenitor of the GRB (e.g., collapsars, neutron star mergers, magnetars...) produces a certain level of gravitational radiation \cite{Kobayashi03,Corsi09,Suwa09}. The background spectrum created should broadly differ from the one predicted here for UHECRs. Moreover, the parameters of the likely acceleration site in GRBs, the shock waves due to the explosion \cite{W01}, are not directly probed by gravitational waves from the central engine. The gravitational radiation emitted by GRB progenitors might therefore not be connected to the capability of GRBs to accelerate UHECRs. 

There have been suggestions that magnetars could be the progenitors of long GRBs \cite{Usov92,Duncan92,Bucciantini09,Metzger10}. If this were the case, whether the sources of UHECRs are GRBs or magnetars would no longer be an issue, but whether the acceleration happens by unipolar induction in the neutron star wind or by Fermi-type acceleration in the GRB shock waves would remain an open question, which could be investigated by the observation of gravitational waves. As we saw, the unipolar induction model requires the magnetar to have a certain level of pulsation and magnetic fields that should naturally lead to the emission of gravitational waves. The detection of gravitational waves with the spectra that we predict here could be one evidence that high enough values for $\mu$ and $\Omega_{\rm i}$ are reached and that unipolar induction acceleration should consequently take place. On the other hand, as mentioned before, the gamma-ray {\it burst} in itself should only lead to a weak signal. 

Another secondary signature of UHECR acceleration in newly born magnetars that could also help distinguish magnetars from GRB scenarios is the neutrino emission. Ref.~\cite{Murase09} calculated that the neutrino flux produced by UHECRs crossing the magnetar surroundings could be detected by IceCube. This signal depends howbeit on the poorly known opacity inside the source (though the column density of the envelope the particles goes through is constrained by the fact that UHECRs need to escape, see Ref.~\cite{Blasi00}). The neutrino spectra found by Ref.~\cite{Murase09} should also be affected by a change in the distribution of magnetar parameters to fit the observed UHECR spectrum. The spectra of cosmogenic neutrinos (produced by the interaction of UHECRs interacting with the radiative cosmic backgrounds while propagating from their source to the Earth) only depends on the overall UHECR spectrum shape, on the source evolution and the injected composition, see e.g., \cite{KAO10} and references therein. The distributions introduced here should thus not lead to new cosmogenic neutrino spectra.\\

We have demonstrated that, by relaxing the assumption of an uniform distribution of magnetar physical parameters, it is possible to reconcile the spectrum of UHECRs produced by these objects with the observed data. The assumptions made on the magnetar population to best fit the observed UHECR spectrum leads to specific gravitational wave background signatures. In some models for which the neutron star deformation by the internal magnetic field is strong and the dipole surface field is a large fraction of the saturation field, the gravitational background signals overlap the sensitivity curves of satellites such as DECIGO or BBO, and are within reach of the sensitivity of the third generation LIGO-type detectors in correlation mode (LIGOIII, Einstein Telescope). The physical assumptions required for such a level of detection can be viewed as optimistic, but are still plausible in the magnetar formation scenarios. The signal obtained can be up to $3-4$ orders of magnitude higher than standard predictions in the frequency range $1-100$~Hz that will be best measured by these future generation instruments. The detection of specific gravitational background spectra as we predict here could probe newly born magnetars as UHECR accelerators, and help solve the long-standing question of the origin of these particles.

\begin{acknowledgments}
I thank Martin Lemoine, Angela Olinto, Kohta Murase, Luc Blanchet and Arieh K\"onigl for very fruitful discussions. This work was supported by  the NSF grant PHY-0758017 at  the University of Chicago, and the Kavli Institute for Cosmological Physics through grant NSF PHY-0551142 and an endowment from the Kavli Foundation.
\end{acknowledgments}

\appendix

\begin{widetext}

\section{Cosmic ray spectrum produced by a population of magnetars \\with a distribution of initial angular velocities $\Omega_{\rm i}$}\label{appendix:Omegai}

For a distribution of initial angular velocities $\Omega_{\rm i}$ with a fixed dipole moment $\mu$ among the stars, one can write:
\begin{equation}\label{eq:dJ}
\frac{\partial J(E,E_{\rm i})}{\partial E_{\rm i}} = W_{\rm geom}\frac{9}{16\pi}\frac{Ic^3}{Ze\mu}\, n_{\rm m} \chi \,E^{-1} T_{\rm loss}(E)  \left(\frac{E_{\rm i}}{E_{\rm i,max}}\right)^{-s} \,\ln\left[ \frac{E_{\rm i}}{E} \frac{1+(E/E_{\rm g})}{1+(E_{\rm i}/E_{\rm g})}\right]\ .
\end{equation}
Integrating over $E_{\rm i}$, one gets:
\begin{eqnarray}
J(E) &=& \int_{E_{\rm i,min}}^{E_{\rm i,max}}\frac{\partial J(E,E_{\rm i})}{\partial E_{\rm i}}\,{\rm d}E_{\rm i}=W_{\rm geom}\frac{9}{16\pi}\frac{Ic^3}{Ze\mu}\, n_{\rm m} \chi\,\left\{
\begin{array}{ll}
{\displaystyle  j_{\Omega,-}(E)} &\quad \mbox{if} \quad E\le E_{\rm i,min}  \\
{\displaystyle  j_{\Omega,+}(E)} &\quad \mbox{if} \quad E> E_{\rm i,min}\, ,
\end{array}\right. \label{eq:jeOmegai}
\end{eqnarray}
where $j_{\Omega,-}(E)$ and $j_{\Omega,+}(E)$ are defined as:
\begin{eqnarray}
j_{\Omega,-}(E) &=&E^{-1} T_{\rm loss}(E)\int_{E_{\rm i,min}}^{E_{\rm i,max}} \ln\left( \frac{E_{\rm i}}{E} \frac{1+E/E_{\rm g}}{1+E_{\rm i}/E_{\rm g}}\right) \left(\frac{E_{\rm i}}{E_{\rm i,max}}\right)^{-s} {\rm d}E_{\rm i}\\
&=& \frac{E^{-1} T_{\rm loss}(E)}{(1-s)^2}\left\{\frac{E_{\rm i}^{1-s}}{E_{\rm i,max}^{-s}}\,\left[(1-s)\ln\left(\frac{E_{\rm i}}{E}\frac{E+E_{\rm g}}{E_{\rm g}+E_{\rm i}}\right)-h(E_{\rm i}/E_{\rm g})\right]\right\}_{E_{\rm i,min}}^{E_{\rm i,max}}\, ,  \label{eq:j-}\\
j_{\Omega,+}(E) &=&E^{-1} T_{\rm loss}(E)\int_{E}^{E_{\rm i,max}} \ln\left(\frac{E_{\rm i}}{E}\frac{E+E_{\rm g}}{E_{\rm g}+E_{\rm i}}\right) \left(\frac{E_{\rm i}}{E_{\rm i,max}}\right)^{-s} {\rm d}E_{\rm i}\\
&=& \frac{E^{-1}T_{\rm loss}(E)}{(1-s)^2}\left\{\frac{E^{1-s}}{E_{\rm i,max}^{-s}}h(E/E_{\rm g})+ E_{\rm i,max}\left[(1-s)\ln\left(\frac{E_{\rm i,max}}{E}\frac{E+E_{\rm g}}{E_{\rm g}+E_{\rm i,max}}\right)-h(E_{\rm i,max}/E_{\rm g})\right]\right\}\, .\label{eq:j+}
\end{eqnarray}
The hypergeometric function is noted:
\begin{equation}
h(x) \equiv \,_2F_1(1,1-s,2-s,-x)\, .
\end{equation}

\section{Cosmic ray spectrum produced by a population of magnetars\\ with a distribution of dipole moment $\mu$}\label{appendix:mu}

We calculate the spectrum obtained for a distribution of dipole moments, when $\Omega_{\rm i}$ is fixed among the stars. Using Eq.~(\ref{eq:eps}), the critical gravitational energy can be expressed as a function of $E_{\rm i}$ as follows: 
\begin{equation}
E_{\rm g} = \frac{\epsilon_{\rm g}^2}{E_{\rm i}}\quad \mbox{with} \quad \epsilon_{\rm g}\equiv\left(\frac{5\,G}{72}\right)^{1/2}\frac{q\eta I}{\beta R^2 c}\Omega_{\rm i}\ .
\end{equation}
One can then write (using Eq.~\ref{eq:E_Omega}):
\begin{equation}
\frac{\partial J(E,E_{\rm i})}{\partial E_{\rm i}} = W_{\rm geom}\frac{9}{16\pi}Ic\,\Omega_{\rm i}^2 n_{\rm m} \frac{\chi}{E_{\rm i,max}} \,E^{-1} T_{\rm loss}(E)  \left(\frac{E_{\rm i}}{E_{\rm i,max}}\right)^{-s-1} \,\ln\left( \frac{E_{\rm i}}{E} \frac{1+EE_{\rm i}/\epsilon_{\rm g}^2}{1+E_{\rm i}^2/\epsilon_{\rm g}^2}\right)\ .
\end{equation}
The integration leads to (we note $S=s+1$):
\begin{eqnarray}
J(E) &=&W_{\rm geom}\frac{9}{16\pi}{Ic}\,\Omega_{\rm i}^2 n_{\rm m} \frac{\chi}{E_{\rm i,max}} \,\left\{
\begin{array}{ll}
{\displaystyle  j_{\mu,-}(E)} &\quad \mbox{if} \quad E\le E_{\rm i,min}  \\
{\displaystyle  j_{\mu,+}(E)} &\quad \mbox{if} \quad E> E_{\rm i,min}\, ,
\end{array}\right. \label{eq:jemu}
\end{eqnarray}
where $j_{\mu,-}(E)$ and $j_{\mu,+}(E)$ are defined (we noted $S=s+1$):
\begin{eqnarray}
j_{\mu,-}(E) &=&
\frac{E^{-1} T_{\rm loss}(E)}{(1-S)^2}\left\{\frac{E_{\rm i}^{1-S}}{E_{\rm i,max}^{-S}}\,\left[(1-S)\ln\left(\frac{E_{\rm i}}{E}\frac{\epsilon_{\rm g}^2+EE_{\rm i}}{\epsilon_{\rm g}^2+E_{\rm i}^2}\right)+h(EE_{\rm i}/\epsilon_{\rm g}^2)-2h'(E_{\rm i}^2/\epsilon_{\rm g}^2))\right]\right\}_{E_{\rm i,min}}^{E_{\rm i,max}}\, , \\
j_{\mu,+}(E) 
&=&\frac{E^{-1} T_{\rm loss}(E)}{(1-S)^2}\left\{\frac{E_{\rm i}^{1-S}}{E_{\rm i,max}^{-S}}\,\left[(1-S)\ln\left(\frac{E_{\rm i}}{E}\frac{\epsilon_{\rm g}^2+EE_{\rm i}}{\epsilon_{\rm g}^2+E_{\rm i}^2}\right)+h(EE_{\rm i}/\epsilon_{\rm g}^2)-2h'(E_{\rm i}^2/\epsilon_{\rm g}^2))\right]\right\}_{E}^{E_{\rm i,max}}\, . \\
\end{eqnarray}
The hypergeometric function $h'$ is defined:
\begin{equation}
h'(x) \equiv \,_2F_1\left(1,\frac{1-s}{2},\frac{3-s}{2},-x\right)\, .
\end{equation}

\section{Cosmic ray spectrum produced by a population of magnetars\\ with a distribution of dipole moment $\mu=f(\Omega_{\rm i}$)}\label{appendix:muOmegai}

We assume a distribution of dipole moments with $\mu=f(\Omega_{\rm i})$ defined in Eq.~(\ref{eq:B_Omega}). For simplicity, we assume that $E_{\rm g}$ is fixed among magnetars (see discussion in Section~\ref{subsection:muf}). Plugging in the relation obtained in Eq.~(\ref{eq:Phialpha}) between $\Phi_{\rm i}$ and $\mu$ into Eq.~(\ref{eq:dJ}), one obtains:
\begin{equation}
\frac{\partial J(E,E_{\rm i})}{\partial E_{\rm i}} = W_{\rm geom}\frac{9}{16\pi}\frac{Ic^3}{q}\, n_{\rm m}\chi\,\left(\frac{A}{E_{\rm i,max}}\right)^{1/3}\,E^{-1} T_{\rm loss}(E)  \left(\frac{E_{\rm i}}{E_{\rm i,max}}\right)^{-s-1/3} \,\ln\left[ \frac{E_{\rm i}}{E} \frac{1+(E/E_{\rm g})}{1+(E_{\rm i}/E_{\rm g})}\right]\ ,
\end{equation}
where 
\begin{equation}
A\equiv{q\eta}\frac{4\pi^{2}}{\alpha^2 R_*^6 c^2}\, .
\end{equation}
The integration is the same as in Appendix~\ref{appendix:Omegai}, replacing the index by $S'=s+1/3$, which leads to:
\begin{eqnarray}
J(E) &=&W_{\rm geom}\frac{9}{16\pi}\frac{Ic^3}{q}\, n_{\rm m}\chi\,\left(\frac{A}{E_{\rm i,max}}\right)^{1/3} \,\left\{
\begin{array}{ll}
{\displaystyle  j_{\Omega,-,S'}(E)} &\quad \mbox{if} \quad E\le E_{\rm i,min}  \\
{\displaystyle  j_{\Omega,+,S'}(E)} &\quad \mbox{if} \quad E> E_{\rm i,min}\, ,
\end{array}\right. 
\end{eqnarray}
where $j_{\Omega,-,S'}(E)$ and $j_{\Omega,+,S'}(E)$ are the functions defined in Eqs.~(\ref{eq:j-},\ref{eq:j+}) where we replace $s$ by $S'$.

\section{Gravitational stochastic background spectra for various magnetar populations}\label{appendix:gw}

\subsection{Distribution of $\Omega_{\rm i}$ (thus of $\nu_{\rm} i$)}
Assuming that the magnetic dipole moment $\mu$ does not vary from one source to another (i.e. $B_{\rm d}$ remains constant), one can re-write the distribution of initial voltages (Eq.~\ref{eq:nm_dE}) as a function of the initial frequency $\nu_{\rm i}$ as follows:
\begin{eqnarray}
\frac{{\rm d} n_{\rm m}}{{\rm d}\nu_{\rm i}} =\frac{{\rm d} n_{\rm m}}{{\rm d}E_{\rm i}} \frac{{\rm d} E_{\rm i}}{{\rm d}\nu_{\rm i}} & = &n_{\rm m}\chi\frac{2 q\eta\mu\pi^2}{c^2}\,\nu_{\rm i}\left(\frac{\nu_{\rm i}}{\nu_{\rm i,max}}\right)^{-2s}\, .
\end{eqnarray}

Taking into account the distribution of sources according to the initial voltage, i.e. to the initial frequency $\nu_{\rm i}$ under our hypothesis, Eq.~(\ref{eq:Ogw}) can be expressed
\begin{eqnarray}
\Omega_{\rm gw}(\nu_0)  &=& 5.7\times 10^{-56} \left(\frac{0.7}{h_0}\right)^2 n_{\rm m,0}\,\kappa\,\nu_0\,\nu_{\rm i,max}^{2s}\, \int_{\nu_{\rm i,min}}^{\nu_{\rm i,max}}\nu_{\rm i}^{1-2s}\,{\rm d}\nu_{\rm i} \,\int_0^{z_{\rm sup}(\nu_{\rm i})} \,{\rm d}z\, 
\frac{R_{\rm SFR}(z)}{(1+z)^2\Omega(z)}\,  \frac{{\rm d}E_{\rm gw}}{{\rm d}\nu}[\nu_0(1+z)]\, \label{eq:Ogwmax}
\end{eqnarray}
with
\begin{equation}
\kappa \equiv \chi\frac{2 q\eta\mu\pi^2}{c^2}\, .
\end{equation}

\subsection{Distribution of $\mu$} \label{subsection:gwmu}
We assume that all magnetars have the same initial frequency $\nu_{\rm i}=\nu_{\rm max}$, and that $\mu$ (thus $B_{\rm d}$) varies from source to source. We can then write:
\begin{eqnarray}
\frac{{\rm d} n_{\rm m}}{{\rm d}\mu} =\frac{{\rm d} n_{\rm m}}{{\rm d}\Phi_{\rm i}} \frac{{\rm d} \Phi_{\rm i}}{{\rm d}\mu} & = &n_{\rm m}\chi\frac{q\eta\pi^2}{c^2}\,\nu_{\rm max}^2\left(\frac{\mu}{\mu_{\rm max}}\right)^{-s} .\label{eq:n_nu}
\end{eqnarray}
And thus:
\begin{eqnarray}
\Omega_{\rm gw}(\nu_0)  = 5.7\times 10^{-56} \left(\frac{0.7}{h_0}\right)^2 \,n_{\rm m,0}\frac{\chi q\eta\pi^2\nu_{\rm max}^2}{c^2}\,\nu_0\, \int_0^{z_{\rm sup}} \,{\rm d}z\, \frac{R_{\rm SFR}(z)}{(1+z)^2\Omega(z)}  \times\nonumber\\\,  \int_{\mu_{\rm min}}^{\mu_{\rm max}}\,{\rm d}\mu\,\left(\frac{\mu}{\mu_{\rm max}}\right)^{-s}\, \frac{{\rm d}E_{\rm gw}}{{\rm d}\nu}[\nu_0(1+z),\mu]\, .\label{eq:OgwB}
\end{eqnarray}
The analytical integration of the last integral over $\mu$ leads to:
\begin{eqnarray}\label{eq:int_mu}
 \int_{\mu_{\rm min}}^{\mu_{\rm max}}\,{\rm d}\mu\,\left(\frac{\mu}{\mu_{\rm max}}\right)^{-s}\, \frac{{\rm d}E_{\rm gw}}{{\rm d}\nu}(\nu,\mu) = \left\{ \frac{\pi^2I}{s-1}\nu \mu\left(\frac{\mu}{\mu_{\rm max}}\right)^{-s}\left[ h'\left(\frac{K\nu^2\mu^2}{\pi^2I}\right)-1\right] \right\}_{\mu_{\rm min}}^{\mu_{\rm max}}\ ,
\end{eqnarray}
where $h'$ is the hypergeometric function defined in Appendix~\ref{appendix:mu}.

\subsection{Distribution of $\mu = f(\Omega_{\rm i})$, i.e., $B_{\rm d}=f'(\nu_{\rm i})$}
Under the assumption of Eq.~(\ref{eq:B_Omega}) connecting the dipole moment $\mu$ to the initial angular velocity $\Omega_{\rm i}$, the magnetar initial voltage can be expressed 
\begin{equation}
\Phi_{\rm i} = \frac{\pi^2\mu\nu_{\rm i}^2}{c^2} = \frac{\pi^2\alpha R_*^3}{2c^2}\nu_{\rm i}^3\, .
\end{equation}
As a consequence, the distribution of magnetars according to the initial frequency reads:
\begin{eqnarray}
\frac{{\rm d} n_{\rm m}}{{\rm d}\nu_{\rm i}} &=& \frac{{\rm d} n_{\rm m}}{{\rm d}\Phi_{\rm i}} \frac{\partial \Phi_{\rm i}}{\partial\nu_{\rm i}} + \frac{{\rm d} n_{\rm m}}{{\rm d}\Phi_{\rm i}} \frac{\partial \Phi_{\rm i}}{\partial\mu_{\rm i}}\frac{{\rm d} \mu_{\rm i}}{{\rm d}\nu_{\rm i}}  \\
& = & n_{\rm m}\chi\frac{3 q\eta\pi^2}{c^2}\,\frac{\alpha R_*^3}{2}\,\nu_{\rm i}^2\,\left(\frac{\nu_{\rm i}}{\nu_{\rm i,max}}\right)^{-3s}\, .
\end{eqnarray}
This yields the gravitational wave spectrum:
\begin{eqnarray}
\Omega_{\rm gw}(\nu_0)  &=& 5.7\times 10^{-56} \left(\frac{0.7}{h_0}\right)^2 n_{\rm m,0}\,\chi\frac{3 q\eta\pi^2}{c^2}\,\frac{\alpha R_*^3}{2}\,\nu_0\,\int_{\nu_{\rm i,min}}^{\nu_{\rm i,max}}\nu_{\rm i}^2\left(\frac{\nu_{\rm i}}{\nu_{\rm i,max}}\right)^{-3s}\,{\rm d}\nu_{\rm i}\,\times\nonumber\\
&&\int_0^{z_{\rm sup}(\nu_{\rm i})} \,{\rm d}z\, \frac{R_{\rm SFR}(z)}{(1+z)^2\Omega(z)}\,  \frac{{\rm d}E_{\rm gw}}{{\rm d}\nu}[\nu_0(1+z),\nu_{\rm i}]\, ,
\end{eqnarray}
with 
\begin{eqnarray}
K&=& \frac{18\pi^4\beta^2R_*^{10}\alpha^2}{5c^2GI}\nu_{\rm i}^2\, .\label{eq:Knui}\\
\end{eqnarray}

\end{widetext}

\bibliography{K11}

\end{document}